\documentstyle[aasms4]{article}
\input{psfig}
 
\lefthead{Burkert \&\ Bodenheimer}
\righthead{Turbulent Molecular Cloud Cores: Rotational Properties}

\begin{document}
 
\def\sol{$_\odot$}                      
\def\x{{$\times$}}
%
%
%
%
\title{Turbulent Molecular Cloud Cores: Rotational Properties$^\dagger$}  

\author{Andreas Burkert\altaffilmark{1}}
\affil{$^1$Max-Planck-Institut f\"ur Astronomie, K\"onigstuhl 17, D--69117 
Heidelberg, Germany; burkert@mpia-hd.mpg.de}

\author{Peter Bodenheimer\altaffilmark{2}}
\affil{$^2$University of California Observatories/Lick Observatory,
Board of Studies in Astronomy and Astrophysics,
University of California, Santa Cruz, CA 95064;
peter@ucolick.org}

\begin{abstract}
The rotational properties of numerical models of centrally condensed,
turbulent molecular cloud cores
with velocity fields that are characterized by Gaussian random fields
are investigated. It is shown that the observed line width -- 
size relationship can be reproduced if the velocity power spectrum
is a power-law with $P(k) \propto  k^{n}$ and $n = -3$ to $-4$. 
The line-of-sight velocity maps of these cores show velocity
gradients that can be interpreted as rotation. For $n = -4$,
the deduced values of angular 
velocity $\Omega$ = 1.6 km s$^{-1}$ pc$^{-1} \times $ (R/0.1 pc)$^{-0.5}$
and the scaling relations between $\Omega$ and the
core radius  $R$ are in very good agreement with the observations.
As a result of the dominance of long wavelength modes, the cores also have a net specific
angular momentum with an average value
of $J/M$ = 7 $\times 10^{20} \times$ ($R$/0.1 pc)$^{1.5}$ cm$^2$ s$^{-1}$ 
with a large spread. Their internal dimensionless rotational
parameter is
$\beta \approx 0.03$, independent of the scale radius $R$.
In general, the line-of-sight velocity gradient of an individual 
turbulent core does not provide a good estimate of its   internal specific
angular momentum. We find however that the distribution of the specific angular momenta
of a large sample of cores  which are described by the same power spectrum
can be determined very accurately from the distribution of their  line-of-sight velocity 
gradients $\Omega$ using the simple formula $j=p \Omega R^2$ where p depends
on the density distribution of the core and has to be determined from 
a Monte-Carlo study. Our results show that for centrally condensed cores
the intrinsic angular momentum is overestimated by a factor of 2-3 if $p=0.4$
is used.
\end{abstract}
           
\noindent 

\section{Introduction: Rotating Cloud Cores}

Although the rotation in the dense ($n \sim 10^4- 10^5$ cm$^{-3}$) cores
of molecular clouds has small dynamical effects compared with gravity, 
it  has important consequences once a core collapses
to form a single star or binary system with associated disks. The 
distribution of separations of binary systems, the distribution of 
disk sizes, and the properties of emerging planetary systems all 
depend on the range of angular momenta among the different cores as well
as on  the angular momentum distributions within individual cores. 
Most theoretical calculations of the collapse
of rotating cloud cores (see Bodenheimer et al. 2000 for a review) 
assume as an initial condition that the core is uniformly rotating; 
furthermore, the observational determination of rotational properties
of cores are based upon a model of uniform rotation (Goodman et al. 1993; 
Goldsmith \& Arquilla 1985; Menten et al. 1984.)  However the material
in molecular clouds is observed to have supersonic line widths over
a wide range of scales indicating a supersonic, irregular velocity field. 
The line width correlates with size, 
providing evidence that has been interpreted in terms of 
turbulent motions (Larson 1981, Myers \& Gammie 1999; see below), probably 
associated with a magnetic field (Arons \& Max 1975). Observed line profiles in
molecular clouds have been shown to be consistent with  Gaussian 
velocity fields with a Kolmogorov spectrum 
(Dubinski, Narayan, \& Phillips 1995, Klessen 2000). 
Even cores on scales of 0.1 pc or less
show non-thermal motions whose velocity dispersion is comparable to, 
but definitely less than, the sound speed (Barranco \& Goodman 1998).
Thus the rotational properties of cores may be more complicated than
the simple law of uniform rotation would indicate. 

The evidence for rotation in the cores of molecular clouds 
(Myers \& Benson 1983; Goldsmith \& Arquilla 1985) consists of  observations
of gradients in the line-of-sight velocity along cuts across the cores.
Goodman et al. (1993; updated by Barranco \& Goodman 1998) 
have observed cores in the size range 0.06 to 0.6 pc
in the NH$_3$ molecule, finding evidence of rotation in 29 out of 43 cases
studied and finding velocity gradients $\Omega$ in the 
range 0.3 to 3 km s$^{-1}$ pc$^{-1}$
(corresponding to 10$^{-14}$ --10$^{-13}$  s$^{-1}$).
Over this range of scales, $\Omega$ scales roughly as $R^{-0.4}$, and the specific 
angular momentum $j \equiv J/M$ as inferred from $\Omega$
scales roughly as $R^{1.6}$, with  a value of
$j \approx 10^{21}$ cm$^2$ s$^{-1}$ on the smallest scales measured.  
The dimensionless quantity $\beta$, 
defined as the ratio of rotational kinetic energy divided by the absolute value of
the gravitational energy, shows no trend with $R$ and has a mean value of about 0.03
with a large scatter. It is also found that cores tend to have gradients
that are not in the same direction as gradients found on larger scales in
the immediate surroundings (Barranco \& Goodman 1998), an effect which
again suggests the presence of turbulence.  

In this paper, following previous suggestions (Goldsmith \& Arquilla 1985; Goodman et al. 1993;
Dubinski et al. 1995) regarding the connection between turbulence and rotation, 
we investigate
 a possible 
origin of rotation in turbulent cores
and investigate the relationship between their line-of-sight velocity maps and
their intrinsic rotational properties.
Unfortunately, a complete and comprehensive theory of turbulence does still not exist.
We therefore adopt the standard simple approach to describe the velocity field inside such
cores by a Gaussian random field. This model assumes random phase correlations between the different
modes. In order to analyze a large statistical sample we also neglect the coupling
between the density and velocity field, which is justified as the flow 
in observed molecular cloud cores is mildly subsonic.
In subsequent papers we plan to include this coupling and to 
analyze in detail how the rotational properties
change during the evolution and collapse of turbulent cores with initial Gaussian random fields.

Here, we try to answer the question whether uniform rotation is a reasonable assumption for such cores,
i.e. whether the velocity gradients, determined from line-of-sight velocity maps,
in combination with the assumption of rigid body 
rotation, provide a good estimate of the
intrinsic specific angular momenta of turbulent cores.
We show that even if the motions in cores are completely 
random, in many cases systematic velocity gradients in the line-of-sight
components of the velocity are present with values that are in good agreement with
the observations, that the cores can have net intrinsic
angular momenta, and that the line-of-sight velocity gradients provide 
on average a good
estimate of their distribution of specific angular momenta.
In \S 2 we describe how random velocity distributions can be derived that
are consistent with observed line width -- size relations. 
\S 3 outlines how the projected angular velocity $\Omega$ is determined from line-of-sight
velocity maps. \S 4 describes the results of a set of 4000 different realizations
of the turbulent velocity field and shows that the model can explain the typical values 
of $\Omega$, $j$, and $\beta$ observed in molecular cores. That the model also
explains the trends with core size is shown in \S 5. 
\S 6 explores the amount of intrinsic specific angular momentum of turbulent
cores and the relationship between the intrinsic angular momentum and the
projected velocity gradient.  Conclusions are presented in \S 7.

\section{Construction of Models  }

The velocity field $\vec{v}(\vec{x})$ can be characterized by its Fourier modes

\begin{equation}
\vec{v}( \vec{x}) = \frac{1}{(2\pi)^3} Re \left[ \int \vec{\hat{v}}(\vec{k}) e^{i \vec{k} \vec{x}}
d^3k  \right]
\end{equation}

\noindent  Dubinski et al. (1995) show that line-of sight velocity 
profiles in molecular clouds are consistent with a Gaussian random field with 
a  Kolmogorov spectrum $P(k) \propto k^{-11/3}$. The relation between the turbulent
spectrum and the line width - size relation has been discussed by Gammie 
\& Ostriker (1996) and   Myers \&  Gammie (1999); the latter authors  also suggest random relative
phases for the spectral components.
Assuming an isotropic velocity field,
the Fourier components $\vec{\hat{v}}(\vec{k})$ are
completely specified by the power spectrum $P(k) = \langle \vec{\hat{v}}^2(k) \rangle$
where $k=|\vec{k}|$.
The power spectrum will depend on the physical properties of the velocity field 
which characterizes molecular clouds and has to be determined from the observations.
Larson (1981) showed that the observed internal velocity dispersion  $\sigma$ 
of a molecular cloud region is well
correlated with its length scale $\lambda$, following approximately a Kolmogorov law

\begin{equation}
\sigma (\lambda) \sim \lambda^q.
\end{equation}

\noindent with $q \approx 0.38$. Later work  found that line width scales with clump size
roughly according to $\sigma \sim \lambda^{0.5}$ (Leung, Kutner, \& Mead 1982; 
Scoville, Sanders, \& Clemens 1986; Solomon et al. 1987).  Additional studies
(Fuller \& Myers 1992; Caselli \& Myers 1995; Myers \& Fuller 1992) measured
slopes in the range 0.25 to 0.75. 
In a more recent investigation, Goodman et al. (1998) find that the power-law 
slope $q$ depends somewhat on 
$\lambda$ with virtually constant line widths ($q=0$) for $\lambda <$ 0.1 pc and
$q$ = 0.5 for larger cores.
Given equation (2), the power spectrum must also follow a power-law  

\begin{equation}
P(k) \propto k^n   ,
\end{equation}

\noindent if $q>0$ . Its slope
depends on the observed value of $q$, and the relation between $q$ and $n$ 
can be determined through filtering the velocity field by passing over it a volume of characteristic
size $\lambda$ and filtering out waves with $k < 1/\lambda$.
This leads to a  variance

\begin{equation}
\sigma^2({\lambda}) = \langle \vec{v}^2(\vec{x}) \rangle_{\lambda} \sim
 - \int_{1/\lambda}^{\infty} P(k) k^2dk \sim \lambda^{-(n+3)} 
\end{equation}

\noindent Note that the integral in   equation (4) converges only if $n < -3$.
Comparing equation (4) with equation (2), we can determine $n$ from the observed
line width-size relationship:

\begin{equation}
n = -3 - 2 q .
\end{equation}

Typical molecular cloud cores with $q \approx 0.5$ will be 
characterized by a power-law index $n \approx -4$ (see also Myers \& Gammie 1999). 
In the following analysis we will explore
the turbulent origin of rotation of molecular cores with a velocity power 
spectrum $-4 \leq n \leq -3$, which 
seems to cover most of the observed range.
The energy spectrum $E(k)$ which corresponds to a velocity power spectrum
$P(k) \propto k^n$ depends on the dimensionality $d$ of the flow and is given
by (Myers \& Gammie 1999) $E(k) \propto k^{n+d-1}$. With $d=3$ and
 $n \approx -4$
the corresponding energy spectrum is $E(k) \propto k^{-2}$.

The velocity field is calculated numerically on a Cartesian 3-dimensional 
grid with $N$ grid cells in each direction
(Cen 1992). 
Due to the limited resolution we
only include modes with  $\lambda > 2R/N$. This is no severe restriction if $N$ is large enough
($N \geq 16$) as waves with small wavelengths do  not contribute to 
global rotational properties of the cores, which are preferentially 
determined by   waves of $\lambda \sim R$.  The results presented below
use $N=64$ with a cutoff at a wavelength of $\lambda < R/32$. Test 
calculations with larger  $N = 256$ and correspondingly smaller cutoff 
wavelengths show that  $N=64$ gives adequate resolution. After generating the  
3-dimensional velocity field $\vec{v}$ using equation (1) we subtract
the center--of--mass velocity.  We adopt a coordinate system where ($x,z$) defines
the plane of the sky and the $y$ direction is along the line of sight. 
Adopting a density distribution $\rho(\vec{x})$, we now
can generate two-dimensional maps, with $N \times N$ pixels, 
of density weighted averaged line-of-sight velocities $V_{LS}(x,z)$  which can be 
analysed and compared with the spectral line maps of observed molecular cloud cores:

\begin{equation}
V_{LS}(x,z) = \frac{\int \rho(\vec{l}) v_y(\vec{l}) d{\vec{l}}}{\int \rho(\vec{l}) dl}
\end{equation}

\noindent where the integration is done along a given line of sight $\vec{l}$ 
through the entire cube. 
Observations indicate that cores are in general centrally condensed
with roughly Gaussian density distributions  
(Ward-Thompson et al. 1994; Andr\'e et al. 1996; review by Bodenheimer et al.
2000). In the following we will use a spherically 
symmetric density distribution of the form 

\begin{equation}
\rho(r) = \rho_c \times \exp \left(-3 \left( \frac{r}{R_{max}} \right)^2 \right).
\end{equation}

\noindent where $r$ is the distance from the center, and 
$R_{max}$ is the outer radius of the core, at which the density 
is assumed to be a factor 20 smaller than the central value $\rho_c$. 
Additional test calculations with a constant density show that the results 
do not depend critically on 
the specific choice of the density distribution. 
Observed cores are typically analysed within a radius $R$ where the surface density
is above half the maximum value. Here
we scale all physical quantities to a typical molecular core with a radius
$R=0.1$ pc. For a core with a density distribution given by equation (7), 
$R \approx 0.5 R_{max}$.
The data presented by Goodman et al. (1993, their table 1) indicate that these
cores have a wide range of masses. 
A rough average value is $M \approx$ 5 M$_{\odot}$, which we adopt as the typical core mass.
Their typical 1-dimensional velocity dispersions are $\sigma_{1d} \approx$ 0.13 km s$^{-1}$ 
(Goodman et al. 1998);
$\sigma_{1d}$ determines the amplitude $A$ of the power spectrum
 $P(k)=A \times k^n$.

\section{Determination of the projected angular velocity and the specific
angular momentum}

In order to compare our models with observations, we analyze the
line-of-sight
velocity maps $V_{LS}$ (see eq. [6]) using the least-squares method 
proposed by Goodman et al. (1993), which minimizes the difference between 
the observed line-of-sight velocity map and the map expected for a 
rigidly rotating core. To provide an approximate
match to the observations, we include only the inner regions of the cores,
where the surface density is above the half-maximum value. In the case of rigid body
rotation the line--of--sight velocity is given by 

\begin{equation}
V_{LS} = V_0 + \omega_z x - \omega_x z
\end{equation}

\noindent where $\omega_z$ and $\omega_x$ are the $z$ and $x$ coordinates of the
three-dimensional angular velocity vector,
 and $V_0$ is the velocity 
of the center of mass. 
This equation assumes that the rotation axis goes through 
the center of the core, which is the origin of the coordinate system. 
Note that we cannot obtain any information about $\omega_y$ from 
the velocity map.  To determine $\omega_x$ and  $\omega_z$ we minimize
the error $\epsilon$:

\begin{equation}
\epsilon = \sum_i 
(V_0 + \omega_z x_i - \omega_x z_i - V_{LS,i})^2
\end{equation}
\noindent where we sum over all pixels $i$ lying within the inner region
defined above. 
Here $x_i$ and $z_i$ are the $x$ and $z$ coordinates of the $i$th pixel, 
and $V_{LS,i}$ is the 
measured line-of-sight velocity in that  pixel. 
We then  solve the following set of equations: 
\begin{equation}
\frac {\partial \epsilon }{\partial  V_0} =  
\frac {\partial \epsilon} {\partial \omega_z} = 
\frac {\partial \epsilon} {\partial \omega_x} = 0.
\end{equation}
Defining the mean value over the ($x,z$) plane of a quantity $q$ as 
$\langle q \rangle = \frac{1}{K \times K} \sum_{i=1}^{K \times K} q_i$, 
where $q_i$ is its value in the i-th pixel and $K = N/2$, 
 and noting that $\langle x \rangle$ =
$\langle z \rangle$ = $\langle xz \rangle$ = 0, we find
\begin{eqnarray}
V_0 & = & \langle V_{LS} \rangle  = 0 \nonumber \\
\omega_x & = & - \frac {\langle z \cdot V_{LS}  \rangle} {\langle z^2 \rangle} \\
\omega_z & = & \frac {\langle x \cdot V_{LS}  \rangle} {\langle x^2 \rangle} \nonumber 
\end{eqnarray}
It can easily be shown that this solution minimizes the error, namely that
the second derivatives are all positive.  
Given $\omega_x$ and $\omega_z$,
 which again are mean quantities
averaged over the $K~\times~K$ surface, 
 we define  the projected $\Omega = 
(\omega_x^2 + \omega_z^2)^{1/2}$ and we determine the angle of the 
projected rotation axis which is defined by tan $\alpha = \omega_z / \omega_x$.
 As an additional rotational property of the core we determine its
total specific angular momentum 

\begin{equation}
j =\left( \sum m \vec{v}\times \vec{x} \right) / \sum m,
\end{equation}

\noindent where  $m$
is the mass of a cell, and  the sum goes over all  
cells of the three-dimensional grid which are located within the
projected half-maximum region which is used  to determine the
projected $\Omega$.

\section{The projected rotational properties of turbulent cores}

Although the velocity fields drawn from the same $P(k)$ are statistically
equivalent,  each realization results from a different set of random
numbers and therefore is unique. As a result we expect that the 
projected and intrinsic rotational properties of the cores may differ significantly
from one case to the next  and will also change as a function of
the index $n$.  Examples are shown in Figure 1, which 
illustrates 3 different line--of--sight velocity maps for each of the cases 
$ n = -4, -3, {\rm and}  -2$~(from top to bottom).
 The frames in the first column correspond to an
example with relatively high $j$ and high $\Omega$, 
the second, to low $j$ and high $\Omega$, and the third, to both low $j$ and 
$\Omega$. One can clearly see that with increasing $n$ the power on
small scales increases, leading to more substructure and  less systematic
motion.
 In the remainder of the paper, we consider only the values
$n = -3, -4$,  which correspond to the range of observed $q$ values
(eq. [2]) in cores (0 to 0.5, respectively). 

The left panels of Figure 1 show examples of a systematic 
velocity gradient which can easily be interpreted as a rotation.
Even for $n= -3$ (second row of Fig. 1) we find  cases with relatively
well-defined velocity gradients. That there indeed exists a global 
projected velocity gradient that could be interpreted as rigid body rotation
is illustrated in Figure 2, which shows the line-of-sight component of the 
velocity averaged over slits parallel to the projected rotation axis as a 
function of distance from the axis  for five cases
for $n = -4$,
similar to  that of the upper left panel of Figure 1. 
The origin of the velocity gradients are dominant long wavelength
modes with small phase shifts with respect to the core center,
like $v(x)=v_0 \sin (\pi x/R_{max})$. In the inner regions of the projected
velocity maps such a wave would indicate a rigid body rotation
with $\Omega \approx \pi v_0/R_{max}$.  The dashed lines in Figure 2 show
the outer (not observed) parts of the cores where $|v|$ reaches a maximum 
and decreases again.

The probability for observing a certain value of a velocity gradient 
can be determined 
as the frequency with which such values would be found given a large number
of projected velocity maps, 
constructed with different sets of random numbers and projection angles. 
The left panels in 
Figure 3 show the distributions of
 $\Omega$ (upper panel), $j$ (middle panel),  and $\beta$ (lower panel), 
 for both
$n = -4$ and $n = -3$, for a set of 4000 random realizations in each case.
Following Goodman et al. (1993) the parameter $\beta$ and the specific internal
angular momentum $j$ are determined from $\Omega$, adopting rigid body 
rotation and a constant core density:

\begin{eqnarray}
j & = & 0.4 \Omega R^2 \nonumber \\
\beta & = & \frac{\Omega^2R^3}{3 G M}
\end{eqnarray}

\noindent  Here $R$ is the radius of the inner region. 
These assumptions are not justified in typical turbulent and centrally
condensed cores. Whether $j$ as determined from equation (13) does indeed
provide a good estimate for the internal specific angular momentum in 
turbulent cores will be discussed in greater detail in \S 6.

For $n = -4$, corresponding to the standard observed 
line width-size relation,  one can see a broad distribution 
in the values of $\Omega$, with a peak at 1.4 km s$^{-1}$ pc$^{-1}$
and an average dispersion of order 1 km s$^{-1}$ pc$^{-1}$.
As expected, the width of the distribution and the value at the peak
decrease with increasing $n$. 
As shown in the second row of Figure 3, the spread of $\Omega$ leads also to
a large spread in the $j$ values which are determined from 
$\Omega$, with $j$ peaking
at $1.7 \times 10^{21}$ cm$^2$ s$^{-1}$. Finally, in the last row of 
Figure  3, the apparent distribution of the 
dimensionless rotation parameter $\beta$ is shown,
derived from $\Omega$, which peaks at $\beta \approx 0.03$.
The distributions of these three quantities for the case 
$n = -3$ (dashed lines in the left panels of Fig. 3)
peak at values 
1/2 -- 2/3 of
 those found for the case $n = -4$. 
Because of the large spread, the average of 
a large number of observations would be required to determine the average rotational
properties of turbulent cores with similar statistical properties. 
For example, the right-hand panels of Figure 3 show the distribution of mean 
values of $\Omega$, j and $\beta$, with each value being the average over
50 different random realizations. It shows that from a typical observational sample of 50 cores,
the average rotational properties can be determined with an accuracy of order 10\%.

In summary, our simulations lead to the following characteristic rotational 
quantities for turbulent cores with $n$ in the range $-3$ to $-4$: 
$\Omega \approx$  0.5 -- 2.0 km s$^{-1}$ pc$^{-1}$, 
$j \approx$ 0.5 -- 2.5 $\times 10^{21}$ cm$^2$ s$^{-1}$,
and $\beta \approx$ 0.01 -- 0.05, with
a large spread in all quantities. These values are in very good agreement with
the observations (Goodman et al., 1993)  which for cores with radii of 0.1 pc
predict $\Omega \approx$ 1.3 km s$^{-1}$ pc$^{-1}$, $\beta \approx 0.03$
and $j \approx 1.2 \times 10^{21}$ cm$^2$ s$^{-1}$.

\section{Scaling Relations}
 
Turbulent cores, where the relevant units are the radius $R$, mass $M$, 
and velocity dispersion $\sigma$, have a well-defined relationship between
j, $\Omega$, $\beta$,  and size: 

\begin{eqnarray}
\Omega & \propto & \sigma /R \nonumber  \\ 
j & \propto & \sigma R   \\
\beta & \propto & (\Omega^2 R^3)/M \propto \sigma^2 R /M. \nonumber
\end{eqnarray}
\noindent The first of these relations follows from the method of
determining $\Omega$ (eq. [11]) and the other two follow from
equation (13) with the assumption of uniform rotation.
Adopting the standard line width--size relation 
$\sigma \propto R^{0.5}$, we find $\Omega \propto R^{-0.5} $ and 
$j \propto R^{1.5}$. The scaling relation for $\beta$ depends on the 
gravitational energy of the core, so an additional mass--radius relation 
is needed. Observations indicate that the velocity dispersion $\sigma^2
\propto M/R$, leading to $M \propto R^2$, and to  a $\beta$  which is
independent of radius. A similar derivation of the scaling
relations has been presented
by Goodman et al. (1993).  Using these relations and the values for
$\Omega$, $j$ and $\beta$ as derived for 0.1 pc cores,
we now can predict the average values
of rotational properties of cores with different sizes. Figure 4 shows that the
calculated average rotational core properties as function of radius
in the range $-4 \leq n \leq -3$ as well as the predicted spread
in $\beta$, are in very good agreement with the observational data.
In fact, in the size range above 0.1 pc, $n=-4$ agrees much better
than $n = -3$, consistent with the line width--size relation. 

\section{The correlation between projected velocity gradient and
specific angular momentum}

The previous section showed that 
a Gaussian random velocity field with a power
spectrum that is in agreement with the line width--size relationship is a
possible explanation of the rotational properties of molecular cloud 
cores, as inferred from line-of-sight velocity maps.
Using this model, we now investigate the relationship between the observed
projected velocity gradient $\Omega$ and the intrinsic angular momentum $j$ of the
cores. 

It is generally assumed that $\Omega$ provides a good estimate of $j$.
This would certainly be expected in the case of solid body rotation.
There, the main uncertainty is the angle $i$ between the line-of-sight direction
and the rotation axis. For random orientations,
the average value of  $\sin^2i $= 2/3, 
and the mean specific angular momentum  can be determined accurately
for a sample of cores of a given size. 

However, if cores are characterized by Gaussian random fields the situation is much
more complex and the assumption of rigid body rotation is not valid.
Now the line-of-sight velocity field does not provide a good estimate of
the amplitudes and phases of the various 
velocity modes in the perpendicular directions.
This is demonstrated in Figure  5, which plots $\Omega$ versus
$j$ for a large sample of cores, generated with a power-spectrum $P(k) \propto k^{-4}$.
$\Omega$ does not correlate
with $j$.
It might at first seem surprising that cores with small specific angular momentum
can show large projected velocity gradients.
To illustrate this effect let
us consider a very simple velocity field with a dominant long-wavelength mode
in the $x$- and $y$-direction and with zero phase shift with  respect to the center:

\begin{equation}
\vec{v} = v_0 \left( \sin (\pi \frac{x}{R_{max}}) \vec{e}_y + \eta \sin (\pi 
\frac{y}{R_{max}}) \vec{e}_x \right) .
\end{equation}

\noindent $R_{max}$ is the radius of the core and $\vec{e}_x$ and
$\vec{e}_y$ are the unit vectors in the $x$ and $y$ directions, respectively.
The case $\eta = -1$ corresponds to a vortex centered at the origin, while
the case $\eta = +1$ places the origin at the ``saddle point" between four
vortex cells. 
Suppose that the line--of--sight direction is along the $y$-axis. As
$\Omega$ is determined from the velocity field inside a radius $r \leq R_{max}/2$,
where $\sin (\pi x/R_{max}) \approx \pi x/R_{max}$, the measured velocity gradient
will be $\Omega  \approx \pi v_0/R_{max}$, independent of $\eta$. 
The left panel of Figure 6 shows the velocity field
in the case of $\eta = -1$.
The core clearly contains a net angular momentum
$j$ around the $z$-axis and $\Omega$ provides a good estimate of $j$.
However, for $\eta = 1$ (right panel of Fig. 6), 
the net angular momentum is $j=0$ whereas
the value of $\Omega$ has not changed. 

In summary, turbulent cores 
are in general not rigid body rotators. Although they could contain a
net angular momentum, as shown in the previous sections, their complex
velocity field makes it impossible to determine the intrinsic angular 
momentum of a core, given its line-of-sight velocity map. This effect
results partly from the fact that compressional velocity components
introduce line-of-sight velocity gradients that are not related to
rotation. 

The situation is however much more promising if one considers the angular
momentum distribution of a large sample of cores that are all
described by the same power spectrum $P(k)$. As $P$ does not depend on
the direction of $\vec{k}$, a set of maps of  the line-of-sight velocity  
contains much more 
information regarding the internal kinematical properties
of the cores.  
This is shown in Figure 7 which compares the distribution of
specific angular momenta
$N(j_{pred}$) as inferred from the line-of-sight 
velocity gradient $\Omega$ (dashed lines)
with the intrinsic distribution of specific angular momenta $N(j)$ (solid lines)
of cores with exponential (Fig. 7a) or constant (Fig. 7b) density
profiles, and 
with $P(k) \propto k^{-4}$.
The values of 
$j_{pred}$ have  been calculated from $\Omega$
using the equation $j_{pred} = p \Omega R^2$ with
$p = 0.14$ for the exponential density sphere and $p=0.4$ for the constant
density distribution.
The intrinsic specific angular momentum $j$,  and from this $N(j)$, is determined
using the full information of the 3-dimensional velocity field (eq. [12]),
summed over the ``observed" inner region. 
The predicted distribution is in excellent agreement
with the real distribution if $p$ is chosen carefully taking into 
account the underlying density profile. For the constant density case
the required value of $p=0.4$ is actually consistent with the real value
of the moment of inertia. In the centrally condensed core, however, a value
smaller than the actual moment of inertia ($p=0.26$) is required to fit the
actual j-distribution.
The distribution of $j_{pred}$ is slightly wider because this value
is determined from random projections.
Note that the $\sin i $-correction is not required in this case.

\section{Conclusions }

Random Gaussian velocity fields with power spectra $P(k) \propto k^{-3}$
to $k^{-4}$ can reproduce both the observed line width -- size relationship
and the observed projected rotational properties of molecular cloud cores.
They therefore can be used in order to investigate their intrinsic velocity
fields in detail or to generate initial conditions for simulations of core collapse
and single star or binary formation.
We have shown that, due to the dominant large-wavelength modes, these cores contain
a non-zero specific angular momentum of order
$J/M = 7 \times 10^{20} \times (R/0.1 {\rm pc})^{1.5}$ cm$^2$ s$^{-1}$.
As a result of the random nature of the velocity field, cores which are statistically
identical, that is  which are described by the same power spectrum, show a large
spread in their rotational properties, which is in qualitative  agreement with the
large spread in observed binary periods (Duquennoy \& Mayor 1991). 
 However, the
median $j$ for pre-main-sequence and main-sequence binaries is about an order of
magnitude less (Simon 1992) than the value we derive for the cores.

The line--of--sight velocity gradient 
does in general not provide a good estimate of the specific
angular momentum of a given core.  However, on a statistical basis, the distribution of
projected velocity gradients $\Omega$  can reproduce very well the distribution 
of the specific angular momenta $j$, assuming $j = p \Omega R^2$,
where $p$ has to be chosen properly through a Monte Carlo study as presented in this
paper. In general, $p$ seems to be smaller than the actual value for the moment
of inertia in centrally condensed cores. As a result of this effect, the specific
angular momenta of cores are overestimated by roughly a factor of 3 if equation
(13) is used. 

It is somewhat surprising that the shape of the
angular momentum distribution as inferred
from the line-of-sight velocity gradients using 
the simple formula of rigid body rotation is in such a good agreement with that of
the intrinsic angular momentum distribution of turbulent cores.
One possible explanation might be that the fluctuation spectrum is dominated
by the large-scale eddies for the values of $n$ that we consider. The 
effect of projection would normally  result in an expectation value
of $j_{pred}$ smaller than the actual  $j$. This effect may be compensated 
through the effects of compressional modes that would contribute to the 
measured velocity gradient and could explain the similar shape and width of 
the two distributions.  More detailed analysis is necessary to explain this
result and to explain the relationship between the intrinsic moment of inertia
and the value of $p$ required to fit the intrinsic angular momentum distribution
from the distribution of line-of-sight velocity gradients. 

In this paper we assumed that the velocity field is uncorrelated with the
spectral line emission and adopted a simple spherically symmetric
density profile. In reality the velocity field will affect the density
distribution and vice versa.  In the case of supersonic turbulence the 
initially Gaussian velocity field will evolve into a system of shocks. 
Numerical models (e.g. Mac Low et al. 1998, Ostriker et al. 1999)
demonstrate that in this case the initial velocity field does indeed not provide a good
estimate of the typical dynamical state of an evolved turbulent cloud.
However, for the scales which are investigated in the present paper, cloud regions are 
mildly subsonic and no strong shocks are expected to form. 
One therefore might not expect a  strong evolution into a dynamical state 
which is very different with respect to the initial power spectrum.

The interaction and dynamical evolution of the velocity and density field in
turbulent cores and its effect on
line-of-sight velocity maps as well as the core collapse and fragmentation
will be investigated in detail in subsequent papers.

\acknowledgements

This work was supported in part through National Science Foundation 
grant  AST-9618548, 
in part through the Deutsche Forschungsgemeinschaft 
(DFG), and in part through a special NASA astrophysics theory program 
which supports a joint Center for Star Formation Studies at NASA/Ames 
Research Center, UC Berkeley, and UC Santa Cruz. We acknowledge
helpful conversations with Richard Klein, Robert Fisher, and Chris 
McKee at a conference in July, 1999, where we learned of their
work on the collapse of turbulent molecular clouds. 
We also would like to thank
L. Blitz,  A. Goodman, and P. Myers 
for interesting discussions and the referee, Ralf Klessen, for many
important comments. AB thanks the staff of 
Lick Observatory for the hospitality during his visits  and PB thanks the
staff of the Max-Planck-Institut f\"ur Astronomie for the hospitality 
during his visits. 

\vfill\eject

\begin{figure}[p] 
\caption{Maps of the normalized line-of-sight velocity for $n = -4$ ({\it top row}), 
$n = -3$ ({\it center row}), and $n = -2$ ({\it bottom row}) as determined 
from eq. (6). 
In the top row, from left to 
right, the values of $\Omega$ in units of km s$^{-1}$ pc$^{-1}$ and the
intrinsic specific angular momentum
$j$ in units of $10^{21}$ cm$^2$ s$^{-1}$ for cores with radii of 0.1 pc are, 
respectively, (1.9, 0.9), (0.2, 1.0), and (0.4, 0.6). 
In the center row, these quantities are
(0.7, 0.4), (0.06, 0.5), and (0.26, 0.2). In the bottom row, 
these quantities are  (0.16, 0.1), (0.004, 0.1), and (0.06, 0.06). 
Blue areas correspond to positive velocities (toward the observer), red corresponds
to zero velocity, and yellow corresponds to negative velocity. 
Each frame  shows the inner ``observed"  region  with dimensions one-half 
those of the full $N^3$ computational grid. 
}
\end{figure}

\begin{figure}[p] 
\centerline{\psfig{figure=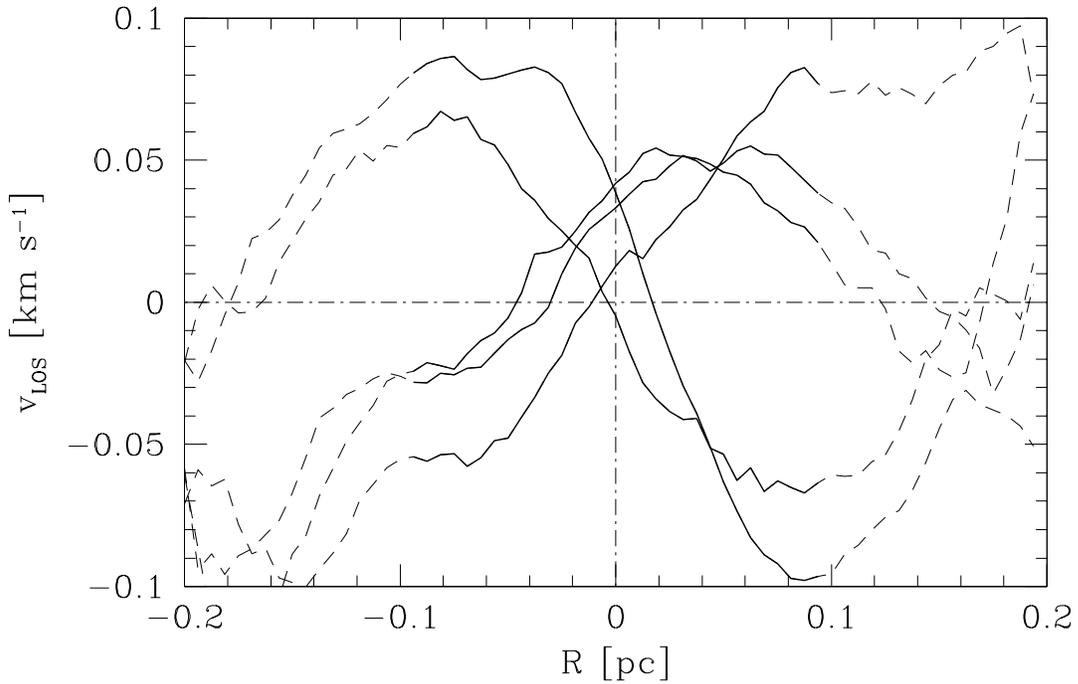,height=16cm}}
\caption{Five randomly chosen examples of simulated cores with large 
projected velocity gradients 
with $n = -4$, 
 similar to the upper left panel of Figure 1. The line-of-sight velocity 
(averaged over a strip parallel to the projected rotation axis with width of 
R/32) is plotted as a function of distance to the projected axis as determined by
the least-squares method.
The dashed lines indicate unobserved regions.
}
\end{figure}

\begin{figure}[p] 
\centerline{\psfig{figure=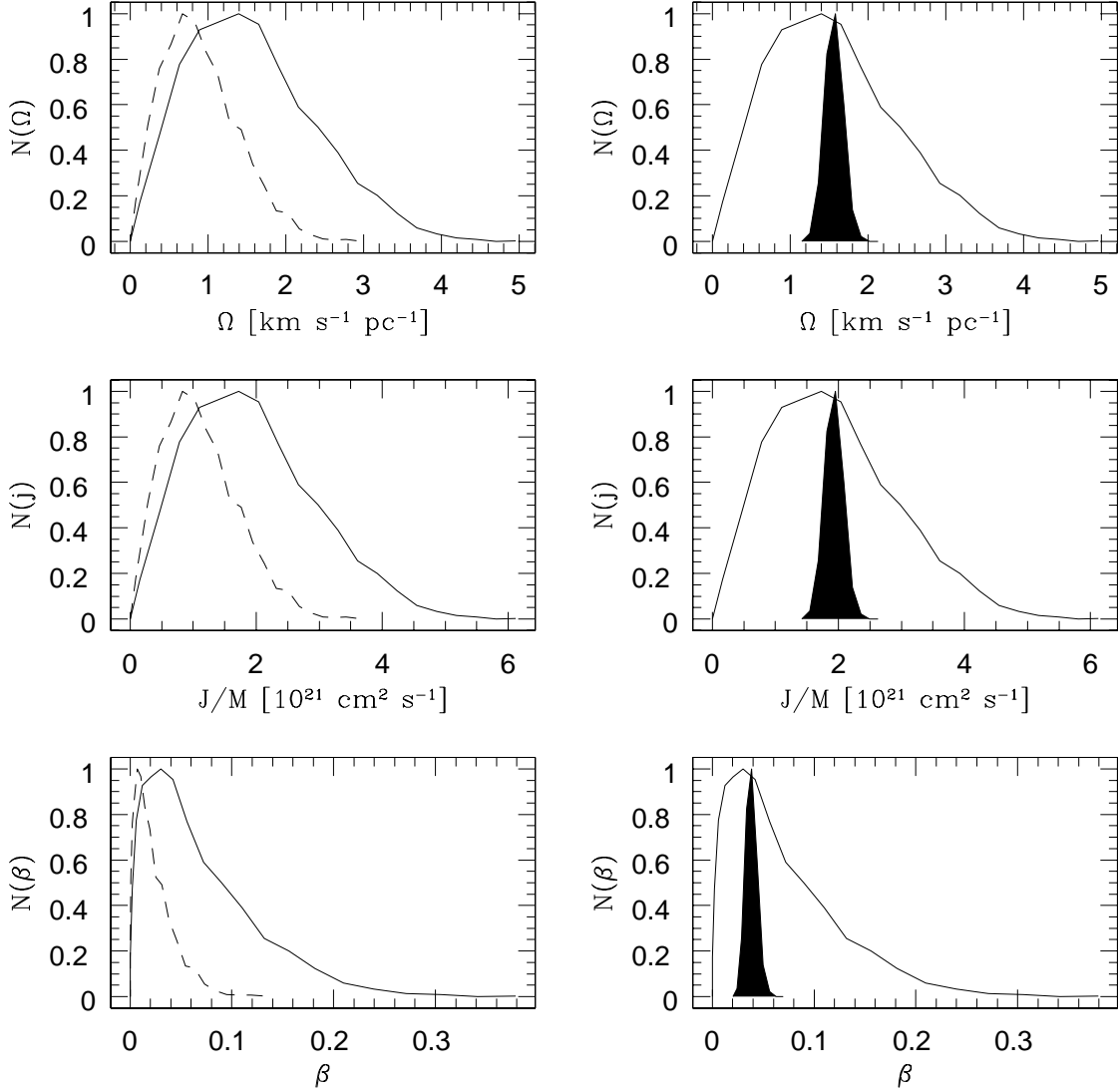,height=16cm}}
\caption{Results of 4000 random realizations of turbulent cores.
Histograms for the projected velocity gradient $\Omega$ ({\it top}), the 
specific angular momentum  $j$ 
as inferred from $\Omega$ ({\it center}) and $\beta$ ({\it bottom})
as inferred from $\Omega$ (see eq. [13]).
In the left panels {\it solid curves} correspond to a power index $n=-4$ and
{\it dashed curves} correspond to $n=-3$. The right panels compare
the histograms for $n= -4$ ({\it solid curves}) with the residual distribution 
with each value being the average over 50 different random  realizations ({\it filled areas}).}
\end{figure}

\begin{figure}[p] 
\centerline{\psfig{figure=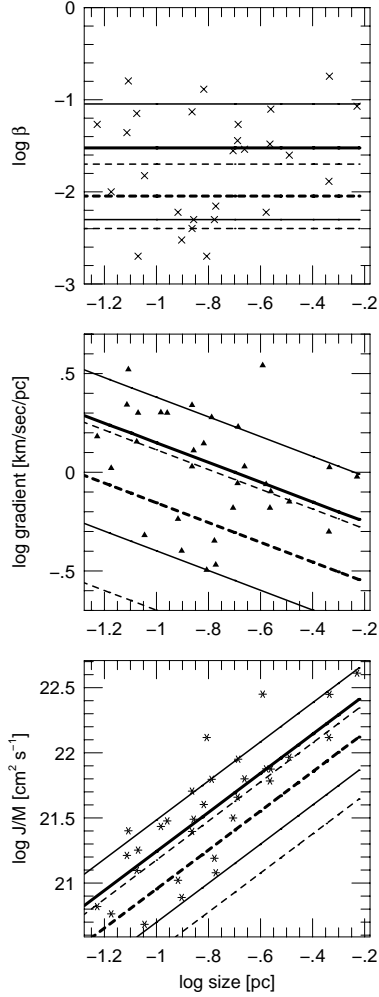,height=16cm}}
\caption{Observed values (Goodman et al. 1993; Barranco \& Goodman 1998) 
of $\beta$ ({\it top, crosses}), velocity gradients 
({\it center, triangles}), and  specific
angular momentum ({\it bottom, stars}) as a function of the size of the core. 
{\it Heavy solid curves:} the trends predicted by the model with $n= -4$.
{\it Heavy dashed curves:} the trends predicted by the model with $n= -3$.
{\it Light solid curves: } the half-maximum points of the calculated 
distribution  for the case $n= -4$.
{\it Light dashed curves: } the half-maximum points of the calculated 
distribution for the case $n= -3$.}
\end{figure}

\begin{figure}[p] 
\centerline{\psfig{figure=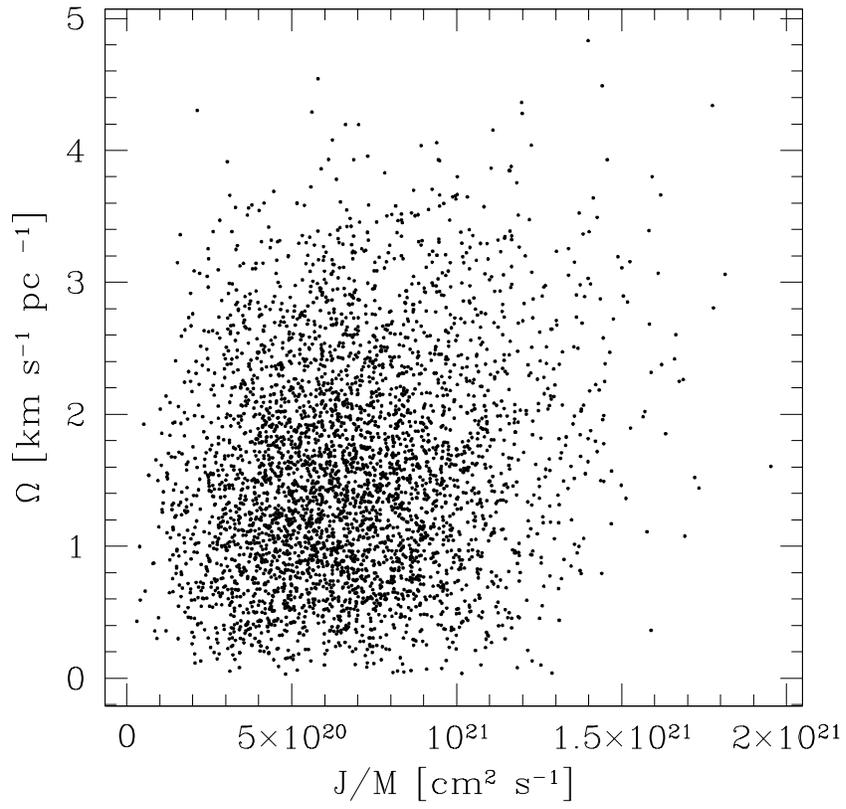,height=16cm}}
\caption{The projected $\Omega$ values of simulated      
cores  ($ n = -4$) 
are  plotted as a function of their internal specific 
angular momentum $j=J/M$. }
\end{figure}

\begin{figure}[p] 
\centerline{\psfig{figure=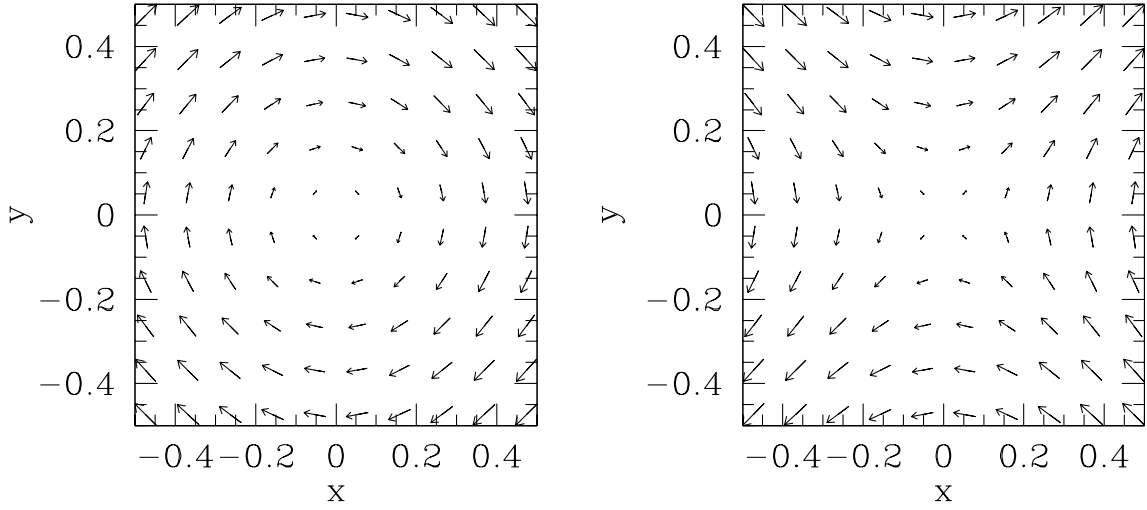,height=16cm}}
\caption{The velocity field which corresponds to eq. (15) is shown with
$\eta = -1$ ({\it left panel}) and $\eta = +1$ ({\it right panel}). The size of the vectors
is linearly proportional to the  absolute value of the velocity.}
\end{figure}

\begin{figure}[p] 
\centerline{\psfig{figure=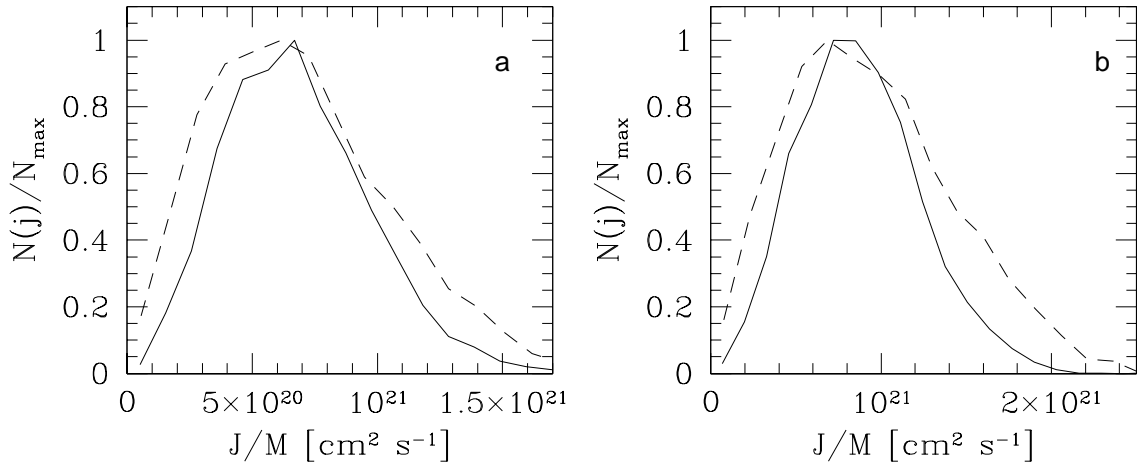,height=16cm}}
\caption{The distribution of specific angular momenta for simulated cores,  as inferred from the projected
velocity gradient determined from the inner region ({\it dashed lines}), is compared with the intrinsic distribution of 
their specific angular
momenta, calculated from the 3-dimensional velocity field ({\it solid lines}).
{\it a):} centrally condensed core. {\it b):} constant density core.}
\end{figure}

\end{document}